\documentclass[twocolumn,preprintnumbers,amsmath,amssymb]{revtex4}

\usepackage{graphicx}
\usepackage{bm}
\begin{document}
\title{A titanium-nitride near-infrared kinetic inductance photon-counting detector and its anomalous electrodynamics}
\begin{abstract}
We demonstrate single-photon counting at 1550 nm with titanium-nitride (TiN) microwave kinetic inductance
detectors. Energy resolution of  0.4 eV and arrival-time resolution
of 1.2 microseconds are achieved. 0-, 1-, 2-photon events are resolved and shown to follow Poisson statistics. We find that the temperature-dependent
frequency shift deviates from the Mattis-Bardeen theory, and the dissipation response shows a shorter decay time than the frequency response at
low temperatures. We suggest that the observed anomalous electrodynamics may be related to quasiparticle traps or subgap states in the disordered TiN films. Finally, the electron density-of-states is derived from the pulse response.
\end{abstract}

\author{J. Gao\footnote{U.S. government work not protected by U.S. copyright.}}
\affiliation{National Institute of Standards and Technology, Boulder, CO 80305}

\author{M. R. Visser}
\affiliation{National Institute of Standards and Technology, Boulder, CO 80305}

\author{M. O. Sandberg}
\affiliation{National Institute of Standards and Technology, Boulder, CO 80305}

\author{F. C. S. da Silva}
\affiliation{National Institute of Standards and Technology, Boulder, CO 80305}

\author{S. W. Nam}
\affiliation{National Institute of Standards and Technology, Boulder, CO 80305}

\author{D. P. Pappas}
\affiliation{National Institute of Standards and Technology, Boulder, CO 80305}

\author{K. D. Irwin}
\affiliation{National Institute of Standards and Technology, Boulder, CO 80305}

\author{D. S. Wisbey}
\affiliation{Department of Physics, Saint Louis University, St. Louis, MO 63103, USA}

\author{E. Langman}
\affiliation{Department of Physics, University of California at Santa Barbara, Santa Barbara, California 93106, USA }

\author{S. R. Meeker}
\affiliation{Department of Physics, University of California at Santa Barbara, Santa Barbara, California 93106, USA }

\author{B. A. Mazin}
\affiliation{Department of Physics, University of California at Santa Barbara, Santa Barbara, California 93106, USA }

\author{H. G. Leduc}
\affiliation{Jet Propulsion Laboratory, California Institute of Technology, Pasadena, California 91109, USA }

\author{J. Zmuidzinas}
\affiliation{Jet Propulsion Laboratory, California Institute of Technology, Pasadena, California 91109, USA }

\date{\today}

\maketitle

Fast and efficient photon-counting detectors at near-infrared wavelengths are in high demand for advanced quantum-optics applications such as quantum key distribution\cite{Hiskett2006} and linear-optics quantum computing\cite{Knill2001}. Superconducting detectors, including superconducting nanowire detectors\cite{Goltsman2001} and transition-edge sensors (TES)\cite{Irwin1995}, show great promise in these applications. For example, TES made from tungsten films have shown over 95~$\%$ of quantum efficiency and photon-number resolving power at 1550~nm\cite{Lita2008}. On the other hand, microwave kinetic inductance detectors (MKIDs) have quickly developed into another major superconducting detector technology for astronomical instruments from submillimeter to x-ray\cite{Day2003,Mazin2009}. The main advantages of MKIDs are that they are simple to fabricate and easy to multiplex into a large detector array.
Recently, the ARCRON camera\cite{Brien2012}, a MKID array made from titanium nitride (TiN) films developed for UV/Optical/NIR imaging and spectroscopy\cite{Mazin2012}, has been successfully demonstrated at the Palomar 200-inch telescope. These TiN MKIDs have already shown good photon-counting and energy-resolving capability, but so far they have been considered only for astronomy applications. In this letter, we describe a single-photon-counting experiment at 1550~nm with TiN MKIDs and discuss their promise for application in quantum optics. Another motivation for the work in this letter is to use these detectors to study the electrodynamics and microwave properties of TiN, which is a relatively new material for MKIDs. Anomalous electrodynamics of TiN is discussed and the density of states $N_0 = 3.9 \times 10^{10}~\mathrm{eV}^{-1}\mu \mathrm{m}^{-3}$ is derived.

MKIDs are thin-film, high-Q superconducting micro-resonators whose resonance frequency $f_r$ and internal quality factor $Q_i$ (or
internal dissipation) change when incoming radiation with photon energy above twice the gap energy ($h\nu > 2\Delta$) breaks Cooper pairs in the superconductor\cite{Day2003}. The measurements of frequency shift and internal dissipation signals are referred to as frequency readout and dissipation readout, respectively. The principle of operation of the MKID, as well as the readout schemes, were explained in detail in a recent review paper\cite{Zmuidzinas2012}.

A recent breakthrough in MKID development is the application of titanium nitride (TiN), a new material for superconducting
resonators that shows a number of ideal properties for MKIDs\cite{Leduc2010}. TiN films have large normal resistivity, high kinetic inductance and low loss. Resonators made from stoichiometric TiN films (Tc $\sim$ 4.5 K) fabricated at NIST and JPL both show $Q_i>10^6$\cite{Leduc2010,Noroozian2012, Vissers2010}. In addition, $T_c$ of sub-stoichiometric TiN films can be adjusted between 0 and 4.5 K by varying the nitrogen concentration during deposition. This tuning allows the gap energy and recombination time to be engineered for a specific application. A lower $T_c$ is desired for photon-counting MKIDs, because the detector responsivity in terms of the frequency shift per incident photon scales as $1/\Delta^2$ \cite{Zmuidzinas2012}.

\begin{figure}[ht]
\includegraphics[scale=1]{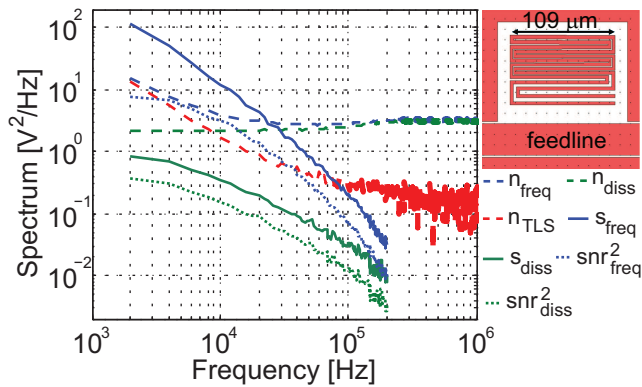}
\caption{(Color online) Spectra of the signal (solid lines), noise (dashed lines) and the signal-to-noise ratio (dotted lines) in the frequency quadrature (blue) and dissipation quadrature (green) measured at 100~mK. The TLS noise (red) is inferred from the difference of the noise in the two quadratures. The resonator shows $f_r = 5.8~$GHz and $Q\sim13,000$ (internal $Q_i\sim100,000$). We use a readout power $P_\mathrm{feedline} \approx -95$~dBm that is 1~dB below resonator saturation (bifurcation).  Inset: the optical lumped-element (OLE) MKID design} \label{fig:noisedesign}
\end{figure}

The device used in this work is made from a 20~nm TiN film with $T_c \sim 1$ K
deposited on a high-resistivity Si wafer ($>10~k\Omega\cdot \mathrm{cm}$). The detector design is adapted from the optical lumped-element (OLE) MKID design used in the ARCONS array\cite{Mazin2012} and is illustrated in the inset of Fig.~\ref{fig:noisedesign}. The inductive part of the resonator, where photons are absorbed and detected, has been specially tapered to give a uniform current distribution and position-independent response. Resonators with design frequency between 4 and 7 GHz are placed directly next to the center strip of a coplanar waveguide (CPW) feedline (without the ground strip in between) to achieve coupling $Q_c\sim 10,000$.

The detector wafer was initially etched by use of Cl-chemistry and showed unexpected high level of frequency noise. It was then re-etched briefly for 3 seconds in SF$_6$-chemistry to clean off the surface residue left by Cl-etch that was recently found to significantly increase the loss and noise in TiN resonators\cite{Sanderg2012}. The detectors are cooled in an adiabatic demagnetization refrigerator (ADR) to 100 mK and measured in the standard MKID readout (homedyne) scheme with a high electron mobility transistor (HEMT) amplifier (noise temperature $T_n\sim 4$ K). A single-mode fiber going into the device box is used to illuminate the entire chip. Outside the ADR, we use a 1550~nm laser diode driven by a function generator to generate optical pulses with width of 200~ns at a repetition rate of 1~kHz. The pulse response signal of the MKID is projected into the frequency and dissipation quadratures, digitized at a sampling rate of 2.5~Ms/s, and processed using a standard Weiner optimal filter, to produce photon-counting statistics.

\begin{figure}[ht]
\center
\includegraphics[scale=1]{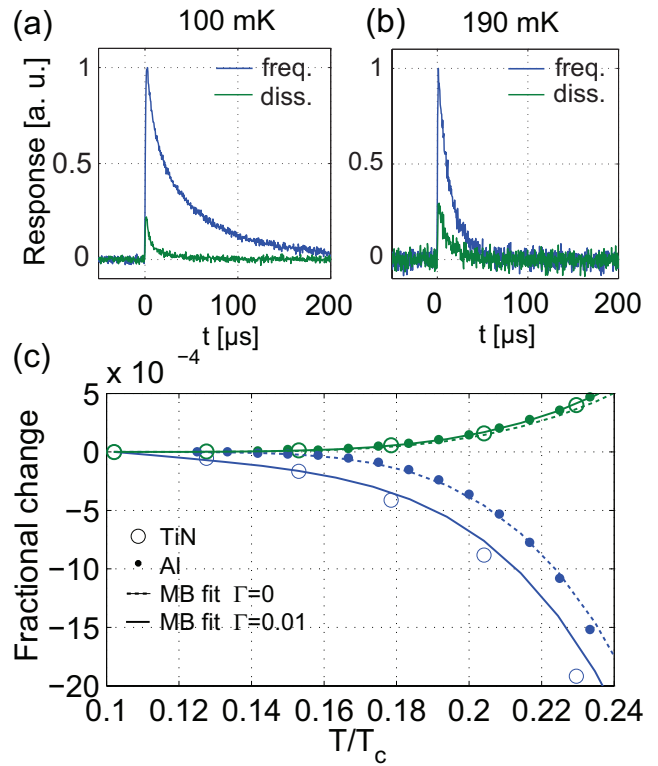}
\caption{(Color online) Anomalous response of TiN MKID. (a) Pulse response at 100~mK. (b) Pulse response at 190~mK (c) Temperature-dependent frequency response $\frac{\delta f_r(T)/f_r}{\alpha\gamma}$ and dissipation response $\frac{\delta 1/2Q(T)}{\alpha\gamma}$ measured for 20~nm TiN ($T_c=1~\mathrm{K}, \alpha=1, \gamma=1$) as compared to 200~nm Al ($T_c=1.2~\mathrm{K}, \alpha=0.276, \gamma=1/3$). $\alpha$ is the measured kinetic inductance fraction \cite{Gao2006} and $\gamma$ is a film-dependent index that takes values of 1 in the thin-film limit and 1/3 in the extreme-anomalous limit\cite{GaoThesis}. Fits using Mattis-Bardeen formula with $\Gamma=0$ and $\Gamma=0.01$ are indicated by the dotted lines and solid lines, respectively. In all the plots, frequency responses are presented in blue and dissipation responses are in green.} \label{fig:anomalodus}
\end{figure}

The detector responses in the frequency quadrature and in the dissipation quadrature measured at 100~mK and averaged over 10000 optical pulses are shown in Fig.~\ref{fig:anomalodus}(a). The pulse (peak) height is about 5 times larger in the frequency quadrature than in the dissipation quadrature, and interestingly, we also see two different pulse decay times, $65~\mu$s for the frequency quadrature and $7~\mu$s for the dissipation quadrature. This anomalous quadrature-dependent response time is discussed below. On the other hand, the measured noise spectra (see Fig.~\ref{fig:noisedesign}) is consistent with the two-level system (TLS) noise picture\cite{Gao2007}: there is $1/f$ excess noise below 10~kHz in the frequency quadrature and no excess noise above the HEMT noise floor in the dissipation quadrature. The signal and noise spectra shown in Fig.~\ref{fig:noisedesign} are used to create the optimal filter. Due to the reduced pulse height and shorter response time, the dissipation readout, although free of excess TLS noise, yields a SNR lower by a factor of $\sim4$ than the frequency readout. Therefore, we used frequency readout for photon counting experiments.

\begin{figure}[ht]
\center
\includegraphics[scale=1]{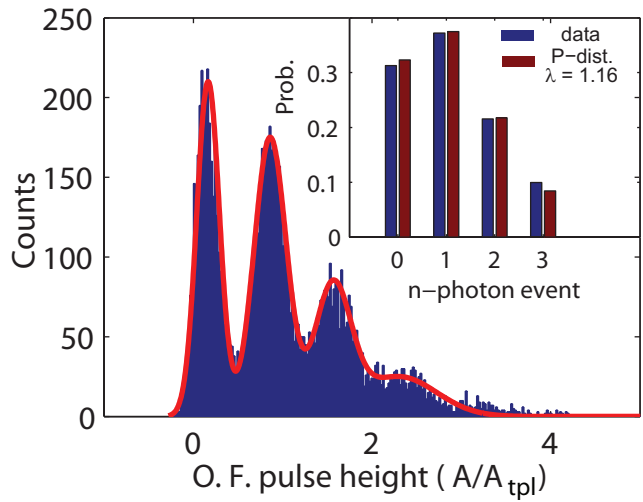}
\caption{(Color online) A histogram of the optimally filtered pulse height (normalized by the template pulse) using the frequency readout. A 4-peak Gaussian fit to the data is shown by the red curve. Inset: the area in each Gaussian peak normalized by the total area (blue) is compared to a Poisson distribution of $\lambda=1.16$ (red).} \label{fig:hist}
\end{figure}

Fig.~\ref{fig:hist} shows a histogram of the optimally filtered pulse height using frequency readout. The first 3 peaks are clearly resolved, and correspond to the events of 0, 1, and 2 photons being absorbed in the detector. We fit the histogram to a model with a superposition of 4 Gaussian peaks, indicated by the red curve in Fig.~\ref{fig:hist}. The FWHM widths of the first 3 Gaussians (uncertainty is large on the 3-photon peak) correspond to the energy resolutions of $\Delta E$ = 0.27~eV, 0.40~eV and 0.44 eV, respectively. The baseline, $\Delta E =0.27~eV$, agrees with a SNR analysis from the signal and noise spectra in Fig.~\ref{fig:noisedesign}. We conclude from the 1-photon peak that our TiN photon counting detector has achieved an energy resolution of $\Delta E = 0.4~$eV, corresponding to an energy-resolving power of $R = E/\Delta E = 2$ at 1550~nm, which is consistent with the energy resolution of TiN MKID demonstrated at UV/Optical wavelengths\cite{Mazin2012} and comparable with the energy resolution of a TES detector operating at the same wavelength\cite{Lita2008}. We have also derived an arrival-time resolution of $\Delta t=1.2~\mu$s from the statistics of the arrival time of the optimally filtered pulses.

It is expected from the stochastic nature of the photon detection process that the n-photon events should obey Poisson statistics. Indeed, the counts in the n-photon peak (proportional to and calculated by the area of each Gaussian) normalized by the total counts match a Poisson distribution of $\lambda=1.16$, as shown in the inset of Fig.~\ref{fig:hist}. In other words, our detector detects an average of 1.16 photons per laser pulsing event in this experiment.


The two different pulse decay times at 100~mK suggest interesting electrodynamic of TiN. We carry out further investigations by repeating the photon counting experiment at elevated bath temperatures between 100~mK and 190~mK. As we raise the bath temperature, we find that the pulse decay time decreases in the frequency response and increases in the dissipation response. At 190~mK, the two quadratures show the same decay times of $16~\mu s$, as shown in Fig.~\ref{fig:anomalodus}(b). To explain this phenomenon, we propose a hypothesis that
low-energy non-dissipative quasiparticle (QP) states exist in TiN that give only a reactive response but no dissipative response.
At lower temperature (100~mK), the non-dissipative QP states are mostly vacant. The high-energy QPs (generated by photons) quickly (within $7~\mu$s) fall into these non-dissipative QP states and then recombine slowly (within $65~\mu$s), leading to an anomalous response with different decay times for the two quadratures. At higher temperature (190~mK), the non-dissipative QP states are mostly filled up with thermal QPs. The high energy QPs directly combine outside these non-dissipative QP states, giving a normal response with the same decay time for the two quadratures. One of the physical candidates for the non-dissipative QP states is quasiparticle traps. TiN films are known to be highly disordered, with localized superconductivity and spatial fluctuations of the gap parameter\cite{Sacepe2008}. In particular, our sub-stoichiometric TiN with $T_c \sim 1~$K sits at a point where $T_c$ is most sensitive to the $N_2$ concentration\cite{Leduc2010}. As a result, we find from DC measurement that $T_c$ is highly non-uniform and varies from 0.4~K to 1.3~K over the entire wafer. It is possible that a locally depressed gap can form a potential well and trap a QP\cite{Sacepe2008}. On one hand, because QPs in the traps are frozen in motion, they do not absorb microwave and can be dissipationless. On the other hand, these QPs are removed (excited) from the Cooper-pair condensate and still contribute to a change in the penetration depth. This is because, for example, the penetration depth is related to the density of superconducting electrons by $\lambda_\mathrm{L} \propto 1/\sqrt{n_s}$ in the London theory\cite{London1935}.

The hypothesis of non-dissipative QP states in TiN is further supported by the temperature-dependent frequency shift and dissipation data that also show anomalous feature (Fig.~\ref{fig:anomalodus}(c)). Compared to Al resonators, the TiN resonator shows excess frequency shift seen as an initial slope in $\delta f(T)$ at low temperatures, which deviates from the prediction of the Mattis-Bardeen's formula. This feature has been reported previously in NbTiN resonators, where a modified MB formula with a gap-broadening parameter $\Gamma$ (Dynes parameter\cite{Dynes1978}) was able to fit the data\cite{Barends2008}. In our case, $\Gamma/\Delta=0.01$ generates a reasonable fit to the data. Further investigation shows that the density of excitation states modified to include the Dynes' parameter has actually introduced sub-gap states near the Fermi level, which play an important role. Excitations to these sub-gap states at $T<\Delta/k$, which was originally forbidden, gives rise to the low-temperature slope in $\delta f(T)$. Interestingly, in this model these sub-gap states also give a much reduced dissipation response. For example, the real part of the complex conductivity $\sigma_1(T)$ calculated from a QP density of states with/without the Dynes' parameter, or with a wide range of $\Gamma$ (as long as $\Gamma/\Delta\ll 1$) shows little difference. Therefore, sub-gap states are another physical candidate for the hypothesis of non-dissipative QP states, and the anomalous temperature and pulse response are consistent under this hypothesis. In fact, sub-gap states deduced from tunneling experiments have been reported for TiN films\cite{Escoffier2004}.

Despite the anomalous response at lower temperatures, we use the pulse response at 190~mK to estimate the single-spin electron density of states, $N_0$, of TiN. This is one of the most important material parameters for MKID design. From the measured average pulse response of $\delta f_r/f_r = 5.6\times 10^{-6}$ and the average photon energy of $E = \lambda h\nu$,  we have derived
\begin{equation}
N_0 = \frac{\eta E S_2}{4 \alpha \gamma \Delta^2 V_\mathrm{sc}\delta f_r/f_r} = 3.9 \times 10^{10}~\mathrm{eV}^{-1}\mu \mathrm{m}^{-3},
\end{equation}
where $\eta$ is the photon-to-QP conversion factor, commonly-assumed to be 0.6, $V_\mathrm{sc} \approx 70~\mu\mathrm{m}^2$ is the volume of the inductive part of the resonator; and $S_2$ is the Mattis-Bardeen factor\cite{Zmuidzinas2012}. This value of $N_0$ is about a factor of 4 larger than the theoretical value previously reported\cite{Leduc2010}.

Our experiment is meant to be a quick demonstration and evaluation of TiN MKIDs for use in quantum optics applications. Many aspects in the experimental configuration and data analysis can be improved. For example, in our current configuration photons are hitting all parts of the resonator without restriction. This may explain the small positive shift of the 0-photon peak seen in Fig.~\ref{fig:hist} and may also contribute to the degradation of energy resolution from the baseline resolution. This can be improved by using a focusing lens or a mask to restrict the light to illuminate only the inductive part of the resonator. Also, no attempt has been made in the current device to improve the optical coupling or the quantum efficiency. In addition to using an antireflection-coating, fabricating the long inductive strip of the MKID on top of an optical wave-guide all on a Si wafer might be an interesting scheme to achieve high coupling efficiency. Our TiN MKID has achieved an energy resolution comparable with that of a TES but is much slower\cite{Lita2008}. To further improve the energy resolution or speed up the detector, a lower-noise amplifier (such as the quantum-limited amplifiers currently under development\cite{EOM2012,CastellanosBeltran2008}) is probably required. This is because the detector noise is already limited by the HEMT noise in both quadratures at the main Fourier component of the pulse ($\sim10$~kHz), as can be seen in Fig.~\ref{fig:noisedesign}.

We thank T. Klapwijk, T. Norguchi, J. Martinis, G. O'Nell and A. Lita for useful discussions. The TiN film was deposited in the JPL Microdevice Lab (MDL), and the device was fabricated in the NIST cleanroom.

\end{document}